**Local Detour Centrality: A Novel Local Centrality Measure for Weighted Networks**

Haim Cohen[1,2], Yinon Nachshon[1], Paz M. Naim[1], Jürgen Jost[4,5], Emil Saucan[6*], Anat Maril[1,3*]

[1]Department of Cognitive Science, The Hebrew University of Jerusalem, Israel

[2] Federmann Center for the Study of Rationality, The Hebrew University of Jerusalem, Israel

[3]Department of Psychology, The Hebrew University of Jerusalem, Israel

[4]Max Planck Institute for Mathematics in the Sciences, Leipzig, Germany

[5]The Santa Fe Institute, Santa Fe, New Mexico, United States of America

[6]Department of Applied Mathematics, ORT Braude College, Karmiel, Israel

Correspondence concerning this article should be addressed to Haim Cohen, The Federmann Center for the Study of Rationality, Givat Ram, The Hebrew University of Jerusalem, Jerusalem 91904, Israel.

email: haim.cohen3@mail.huji.ac.il; phone: +972-54-635-9838

**Abstract**

Centrality, in some sense, captures the extent to which a vertex controls the flow of information in a network. Here, we propose Local Detour Centrality as a novel centrality-based betweenness measure that captures the extent to which a vertex shortens paths between neighboring vertices as compared to alternative paths. After presenting our measure, we demonstrate empirically that it differs from other leading central measures, such as betweenness, degree, closeness, and the number of triangles. Through an empirical case study, we provide a possible interpretation for Local Detour Centrality as a measure that captures the extent to which a word is characterized by contextual diversity within a semantic network. We then examine the relationship between our measure and the accessibility to knowledge stored in memory. To do so, we show that words that occur in several different and distinct contexts are significantly more effective in facilitating the retrieval of subsequent words than are words that lack this contextual diversity. Contextually diverse words themselves, however, are not retrieved significantly faster than non-contextually diverse words. These results were obtained for a serial semantic memory task, where the word's location constitutes a significant mediator in the relationship between the proposed measure and accessibility to knowledge stored in memory.

**Introduction**

Many applications use networks to represent complex interactions between objects. A network is composed of two elements: objects (vertices) and their interactions (edges). The interaction between objects may express, for instance, road traffic (Gao et al. 2019), brain imaging (Elumalai et al. 2022), proteins (Rao et al. 2014), stocks (Sandhu et al. 2016), or semantic relations among words (Nachshon, Cohen, and Maril 2022).

The topology of a network that is constructed by real-world data is non-homogeneous, meaning that the amount of information that travels through each vertex varies from one vertex to another. Many networks, for example, exhibit a scale-free property, which means that they follow a power law distribution (Barabási 2009). In such a case, most of the vertices are connected to only a few other vertices, and a small number of vertices are connected to many other vertices. As a result, a small number of highly connected vertices controls the flow of information in the network. By uncovering patterns in the interactions among vertices, we can identify the vertices with the greatest influence on network stability and robustness. These central vertices may explain phenomena such as the spread of disease among cities or the spread of information in a social network (Das et al. 2018; Grubb et al. 2021).

Intuitive understandings of centrality have generally taken three forms. In the first form, centrality measures the *efficiency* of a vertex in communicating with other vertices; the focus is the central object's interactions with other vertices, with whom the central object is assumed to share information (Freeman 1978). In the second form, a central vertex is characterized by the *feedback* that it receives from its neighbors. The more that its neighbors are themselves central, the greater the centrality of the vertex in question. This is the notion behind eigenvector, PageRank, and Katz centralities. In the third form, centrality is related to betweenness: objects are central in the sense that they stand between other objects and serve as a bridge over which information flows between those objects. This idea is based on the insight that a central vertex *controls* the information that flows in its environment Tutzauer 2007, Kivimäki et al. 2016).

Several measures based on this third understanding of centrality already exist. Shortest path betweenness considers the flow of information along the shortest paths (Freeman 1978), an idea that Freeman et al. (1991) expanded into flow betweenness centrality, which considers non-geodesic paths. The notion was pursued further by Newman (2005) and Brandes and Fleischer (2005), who developed random-walk betweenness and electric current flow betweenness, which considers the contribution of all paths between the source and target in computing the flow of an electric current. Notably, Bubboloni and Gori (2022) proposed a flow betweenness centrality that considers the amount of information that gets lost if it is impossible to pass through a given vertex.

In offering Local Detour Centrality (LDC), our novel centrality-based betweenness measure, we, too, take the third understanding of centrality as our starting point. Consistent with the shortest path approach, LDC is a measure of connectedness constant that indicates the tendency of a given vertex to lie on the shortest possible path between other vertices. The LDC of vertex $v$ indicates whether the possible path between $v_i$ and $v_j, i \neq j$ is significantly shorter when $v$ lies between them than when $v$ does not lie between them. For the sake of simplicity, we compare only the shortest path to its alternatives. In this perspective, the vertex's control of information flow in the graph means that the vertex systematically offers a faster way to move from one vertex to another compared to alternative vertices. The vertex in question does not necessarily offer the shortest possible path; instead, it offers the greatest difference between the short path that passes through it and the best alternative that does not pass through it.

This study presents two distinct innovations. From a general perspective, we offer an alternative centrality measure for weighted networks. This alternative is important to both the directed and the undirected case, since our measure takes into account a new invariant that captures whether or not there are shorter detours for a path along some edge. Second, in our demonstration of the use of LDC in a semantic network, we present a psycholinguistic interpretation of the measure as a means of capturing the contextual diversity (CD) of a word, meaning the extent to which that word occurs in different contexts.

It is possible to identify CD because of the way in which semantic context is expressed in the graph. A semantic context is a set of words that are interconnected by short possible paths; this set is connected by longer possible paths to words that belong to other semantic contexts (Newman 2018). These longer possible paths become shorter, however, when they run through words that belong to several contexts—in other words, to CD words, which bridge the various contexts to which they belong. For example, in a semantic graph in which the vertices are names of animals, we can assume that "wolf" and "cat" belong to two different contexts while "dog" belongs to both contexts. In this case, the sum of the weights from "wolf" to "dog" and then from "dog" to "cat" is shorter than the possible path from "wolf" to another animal and then from that other animal to "cat." In other words, the possible paths that go through "dog" are shorter than the possible paths that go through other words. The word "dog," then, functions as an intermediary between vertices since it binds unrelated words from different semantic contexts. The novel measure presented in this study reflects the extent to which a word like "dog" mediates between any pair of words in its vicinity. The more a word functions as an intermediary—in other words, the more paths that a word shortens on the graph—the greater the number of semantic contexts to which that word belongs.

The purpose of this demonstration is twofold. First, from a general perspective, we investigate whether LDC can be distinguished from existing centrality measures. Second, within the context of semantics, we examine whether LDC can measure CD in a semantic graph. We validate our new measure psychologically as we control for frequency, referring to findings that high CD words are processed faster than low CD words in tasks involving lexical decision, word naming, and recognition (Adelman et al. 2006; Brysbaert and New 2009; Johns et al. 2012; Baayen 2010; Caldwell-Harris 2021; Steyvers and Malmberg 2003; Lohnas et al. 2011). In brief, CD words are associated with greater accessibility. We test this relationship in two ways. First, we examine whether CD words are retrieved more quickly than non-CD words in a semantic memory task. Second, we explore the extent to which a CD word facilitates the retrieval of the words that follow it. We posit that once a CD word has been retrieved, the fact that it belongs to many contexts should make it easier to retrieve the next word.

In sum, we expect to find an effect of CD on the level of accessibility. Here it is worth noting two fundamental differences between our study and the existing literature on CD. Our work focuses on semantic memory in a serial task, while most of the literature deals with episodic memory, and in most cases, a non-serial task. We discuss these differences in greater detail in our concluding section.

The remainder of this paper is organized as follows. After we describe related work on centrality measures in our first section, we devote our second section to an introduction of LDC. The third section presents our case study data, and the fourth section examines how LDC differs from measures that have already been proposed. The fifth section presents LDC as an expression of CD words and examines the relationship between CD and frequency. The sixth section investigates the relation between LDC and accessibility to information stored in memory. The final section offers some conclusions.

**Section 1. Related work**

Each centrality measure that appears in the work related to our study falls into one of the three intuitive understandings of centrality that we have already described. Degree, closeness, and number of triangles capture the extent to which a word is close to other words. PageRank captures feedback centrality. Betweenness and LDC capture the extent to which a word functions as an intermediary. We will examine empirically the extent to which LDC differs from all these other measures, including betweenness, with which LDC shares the same notion of centrality, and then we will examine the robustness of these differences. To do this, we must first introduce the measures.

*Degree*

For a directed graph $G$ and the vertex $v \epsilon V$, the in-degree of vertex $v$ refers to the number of arcs that incident from $v$. The out-degree refers to the number of arcs that incident to $v$.

*Closeness*

Closeness is the inverse of farness, which is defined as the mean of the shortest paths to all other vertices (Borgatti and Everett 2006; Freeman 1978). Closeness can be interpreted as the expected time of arriving at a word through the graph's shortest paths. The gist of this metric is to assign more importance to the vertices that are closest. The definition is as follows:

$$C(v) = \frac{N-1}{\sum_u \delta(v, u)} \qquad (2)$$

where $\delta(v, u)$ represents the shortest path between $v$ and $u$. This measure takes weight into account by averaging the shortest paths that emerge from $v$.

*The Number of Triangles*

The number of triangles calculates the number of undirected 3-cliques for each vertex in the graph. This measure is used to detect vertices that belong to numerous cliques. It is worth noting that this measure is closely related to the clustering coefficient.

*PageRank*

The basic idea of the PageRank algorithm, first introduced in a Google paper (Brin and Page 1998), is that a central vertex is determined not only by the number of incoming edges (in-degree) but also by the level of importance of the incoming vertices. $T$ represents the set of vertices, $N_u$ represents the number of vertices to which vertex $v$ points, and $S_V$ represents the set of vertices pointing to vertex $v$. Finally, $\alpha$ is the damping factor of the probability of jumping from a given vertex to another random vertex in the graph. PageRank is computed as follows:

$$\text{PR(v)} = \sum_{u \in S_v} \frac{PR(u)}{N_u} + \frac{1-\alpha}{T} \tag{3}$$

While LDC may be compared with eigenvector centrality for undirected graphs, here we focus exclusively on directed graphs. The systemic comparison of centrality measures in an undirected graph is a subject for further research.

*Betweenness*

Shortest path betweenness was introduced by Freeman (1977) in order to quantify the extent to which a vertex tends to be on the shortest paths between other vertices—in other words, to serve as an intermediary. Betweenness for a vertex $v$ is defined as follows:

$$\gamma(v) = \sum_{i \neq v \neq j \in V} \frac{\sigma_{ij}(v)}{\sigma_{ij}} \tag{4}$$

where $\sigma_{ij}(v)$ represents the number of shortest paths between $i$ and $j$ that go through $v$, and $\sigma_{ij}$ represents the total number of shortest paths between $i$ and $j$. Brandes and Fleischer (2005) extended this idea to take into account all possible paths, not just the shortest ones, between two vertices.

**Section 2. Local Detour Centrality: definitions and examples**

Let $G = (V, E, w)$ be an edge-weighted and directed graph in which $w(v_i, v_j)$ represents the weight from $v_i$ to $v_j$ and let $\delta(v_i, v_j)$ denote the shortest path from $v_i$ to $v_j$ based on Dijkstra's shortest path algorithm. For any vertex $v$,

a. Let $L = \{v_1, v_2 \ldots v_n\}$ such that any $v_i \in L$ if $\delta(v, v_i) \leq r$ or $\delta(v_i, v) \leq r$. The number $r = \frac{1}{|V(G)|}\sum_{v_i, v_j \in V} \delta(v_i, v_j)$ is called the threshold.

b. Let $G_v \subset G$ be a complete graph with $V(G_v) = L$ where in this case $w(v_i, v_j) = \delta(v_i, v_j)$.

To calculate local intermediateness, for any vertex $v$, let us first introduce some further notations:

c. Let $GN = (V, E, w)$ be the complete graph, where the weights are calculated according to the Dijkstra's shortest path algorithm on the following weights:

$$w'(v_i, v_j) = \begin{cases} \max_{v_k, v_l \in V}(w(v_k, v_l)) & v_j = v \text{ or } v_i = v \\ w(v_i, v_j) & otherwise \end{cases}$$

Since both graphs $G_{\bar{v}}$, $G_v$ have the same edges but not the same weights, we denote E' = $E(G_{\bar{v}})$ = $E(G_v)$, and then the local detour centrality (LDC) of the vertex $v$ is defined as follows:

$$LDC(v) = \frac{1}{|L|} \sum_{e \in E'} \left( G_{\bar{v}_e} - G_{v_e} \right) \quad (1)$$

$G_v$ denotes the shortest paths matrix $(L \times L)$ in the vicinity of $v$ when it is possible to pass through $v$. $G_{\bar{v}}$ denotes the shortest paths matrix $(L \times L)$ where the short paths are constructed by $GN$ in which the distance to and from $v$ receives the maximum value of the graph. Next, we perform an element-wise subtraction of the matrix $G_{\bar{v}}$ from matrix $G_v$ such that the higher the value is, the longer are the shortest paths in $G_{\bar{v}}$ compared to the shortest paths in $G_v$. In such a case, $v$ binds unrelated objects in its local environment.

Vertex $v$ will be considered an intermediary based on a comparison between the shortest paths linking each pair of vertices in the local environment of $v$, given that it is possible to pass through $v$, and the shortest paths linking those vertices, given that it is not possible to pass through $v$. A vertex constitutes a local intermediary if the paths are shorter when they pass through $v$. In Figure 1, for example, the LDC of vertex A in (a) is lower than the LDC of vertex A in (b), even though the weights going in and out of A are shorter in (a) than they are in (b). The LDC is higher in (b) because A more significantly shortens the path from B to C than does the second shortest path between the two points (i.e. B-D-C).

**Figure 1.** *Toy example for Local Detour Centrality*

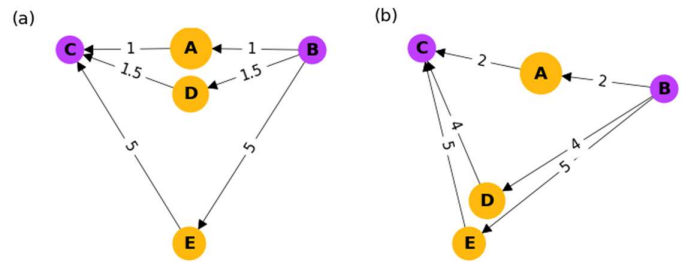

This attention to the weight of the edges constitutes the main distinction between both forms of betweenness centrality and LDC. Neither short path betweenness nor flow betweenness is affected if there is a short detour that bypasses a central node, but LDC becomes much smaller when such a bypass is readily available. While short path betweenness and flow betweenness consider the amount of information that travels through a vertex, LDC considers the extent to which the traffic would have deteriorated had the path through the vertex been interrupted.

An important property that we applied in LDC is *locality*, which figures in a measure we call local intermediateness. This measure reflects to what extent $v$ functions as an intermediary within the group of $R$-neighbors sufficiently close to $v$. The degree of closeness to $v$ is defined by an upper bound, a free parameter that determines the $R$-neighbors of $v$ on which the intermediateness is calculated. For our purposes, we define the $R$-neighbors of $v$ as the set of vertices whose distance from $v$ is smaller than the average distance in the graph.

**Section 3. Case study: LDC as contextual diversity in a semantic network**

In this section, we aim to examine empirically the relationship between the proposed measure and centrality, and we will discuss whether LDC is indeed different from other centrality measures. Then we will demonstrate a possible use of LDC in semantic network. As a first step, we introduce our data and graph construction.

*Data and Tools*

The experiment was conducted on two groups ($N = 2047$), both consisting of native Hebrew speakers. One group was recruited from the Hebrew University of Jerusalem (HUJI: $N = 691$; M:F = 1:1.002; mean age = 24.6 years; range: 18-39); the members of this group received coupons for coffee. The second group (P4A: $N = 1356$ ; M:F = 1:1.96; mean age = 29.07 years; range: 18-40) was recruited through the panel4all website, and participants were compensated with gift certificates from the panel4all organization. The ethics committee of the Department of Psychology at the Hebrew University of Jerusalem approved all experimental procedures.

Participants were given one minute in a category fluency test (CFT); the task was to produce as many unique words as possible within the semantic category of animal names. Participants from HUJI were recorded on a Philips DVT4010, and soundtracks were transcribed with the PRAAT program (Boersma and Weenink 2021), which gave us the words as well as the time signatures for the beginning and end of each word. Participants from P4A were recorded on a phone application, and these soundtracks, too, were transcribed via PRAAT.

Two lists were generated for each participant: a list of words and a list of timestamps, with each timestamp indicating the start time of the word's retrieval. The timestamps start at 0, indicating the beginning of the recording, and end at 60.

All of the Python code, raw data and other supplementary material are publicly available via the Open Science Framework (Cohen 2022).

We will now describe how we construct the semantic graph based on the dataset we have just presented. Let $G = (V, E)$ be an edge-weighted and directed graph, with $V = \{v_1, v_2 \ldots v_n\}$ a set of vertices and $E = \{(v_i, v_j)\}$ a set of edges or links between words $v_i$ and $v_j$ in $V$. The words are the vertices, and the edges reflect the relationship between words. The *R-neighbors* of vertex *v* are the group of vertices at a distance $\leq R$ from *v*. The weight $w(v_i, v_j)$ is based on the assumption that the closer the semantic relationship between two words, the faster the transition between these two words (Collins and Loftus 1975).

The weights of the edges are based on Nachshon's proposal (Nachshon, Cohen, and Maril, 2022) and are calculated by a "distance" function that assumes, as expected for a metric, that the "distance" is non-negative. However, we do not assume symmetry such that $w(v_i, v_j) \neq w(v_j, v_i)$. We also allow violation of the triangle inequality. Thus, our "distances" do not constitute a true metric.

Our "distance" function calculates the weights of the edges as follows. For any ordered pair of vertices $v_i, v_j$, *p = p(s)* is a sublist of participant *s*; the sublist starts at $v_i$ and ends at $v_j$ and denotes the amount of normalized time that it took *s* to traverse from $v_i$ to $v_j$. Here, each timestamp was normalized by the number of words that *s* produced. The "distance" function has two free variables that determine the upper and lower boundaries, WS (window size) and MS (minimum subjects):

1. The upper boundary window size (WS) which is an integer defines the maximum number of words between $v_i$ and $v_j$. The sublist *p* is therefore taken into account when the number of words between $v_i$ and $v_j$ is less than or equal to the number WS.

2. MS is a number defining the lower boundary, which is the minimum number of p's containing $v_i$ and $v_j$ in that order, and with at most WS words between them.

Let $P$ be the set of the amount of times it took any $p$ to traverse from $v_i$ to $v_j$ up to WS words. Then, the weights between the ordered pair $v_i$ and $v_j$ is defined as being the median of the numbers in the set $P$, if $|P|>$MS. Otherwise, there will no distance between $v_i$ and $v_j$ is defined. In other words, a distance is well-defined iff $|P| >$ MS. Given a path on the graph, its length will be the sum of the weights (or distances) of the edges composing it, as defined above.

Note that not every pair of vertices $i,j$ gets a weight, only those pairs that at least MS subject retrieved $i$ and then $j$ when the number of words in-between them is at most WS.

**Section 4. Differentiation**

As we have already pointed out, each centrality measure falls into one of the three ways that centrality has been intuitively understood. Centrality as efficiency is reflected in degree, number of triangles, and closeness; centrality as feedback is reflected in PageRank; and centrality as control is reflected in betweenness and our own measure, LDC. We will examine empirically the extent to which LDC differs from all these other measures, including betweenness, with which LDC shares the same notion of centrality, and then we will examine the robustness of these differences.

To study these questions, we performed a correlation test between every possible pair of centrality measures. All the analyses were performed on the set of vertices that received a value for each centrality measure. Additionally, the correlation between each pair of centrality measures was tested on a range of two parameters of the distance function that we have already introduced. The first parameter, MS (minimum subjects per edge), comprises the values [3,5,7,9,11,13,15,17,19,21]. The second parameter, WS (window size), or the maximum number of words that determine the distance between a given pair of words, comprises the values [1, 2,…,9]. Note that all networks were built by the lists of words that the subjects provided. However, each network has a distinct set of weights since the weights of the edges are determined by the free parameters WS and MS. Words were excluded if one of the two centrality measures assigned a value greater or smaller than 2.5SD from the mean. Figure 2 illustrates the number of vertices on which the correlation test was performed, with each cell representing the number of vertices for a pair of parameters MS and WS. In Figure 3, each matrix represents the correlations between a pair of centrality measures, with the x-axis representing a range of MS and the y-axis representing a range of WS.

**Figure 2.** *Vertices on which the correlation test was performed*

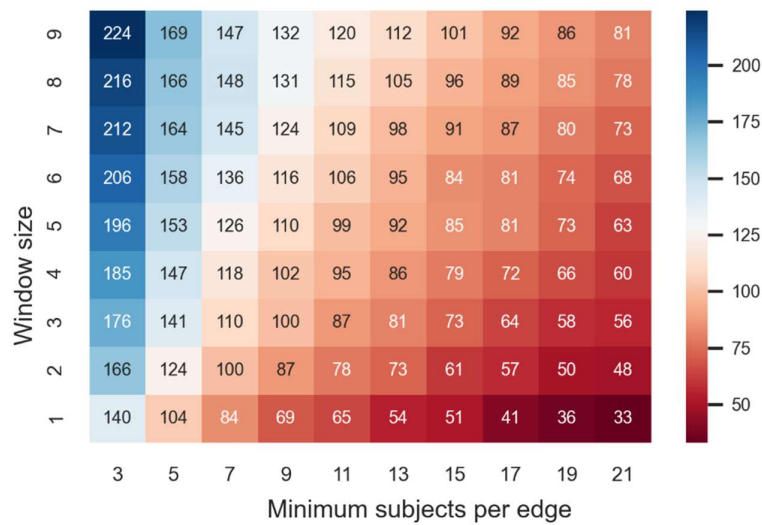

*Legend.* Each cell represents the number of vertices for a pair of parameters MS (minimum subjects per edge) and WS (window size). The x-axis denotes the range of MS values, and the y- axis denotes the range of WS values. The color highlights the number of vertices. As the cells increase in redness, the number of vertices decreases.

**Figure 3.** *Correlations between pairs of centrality measures*

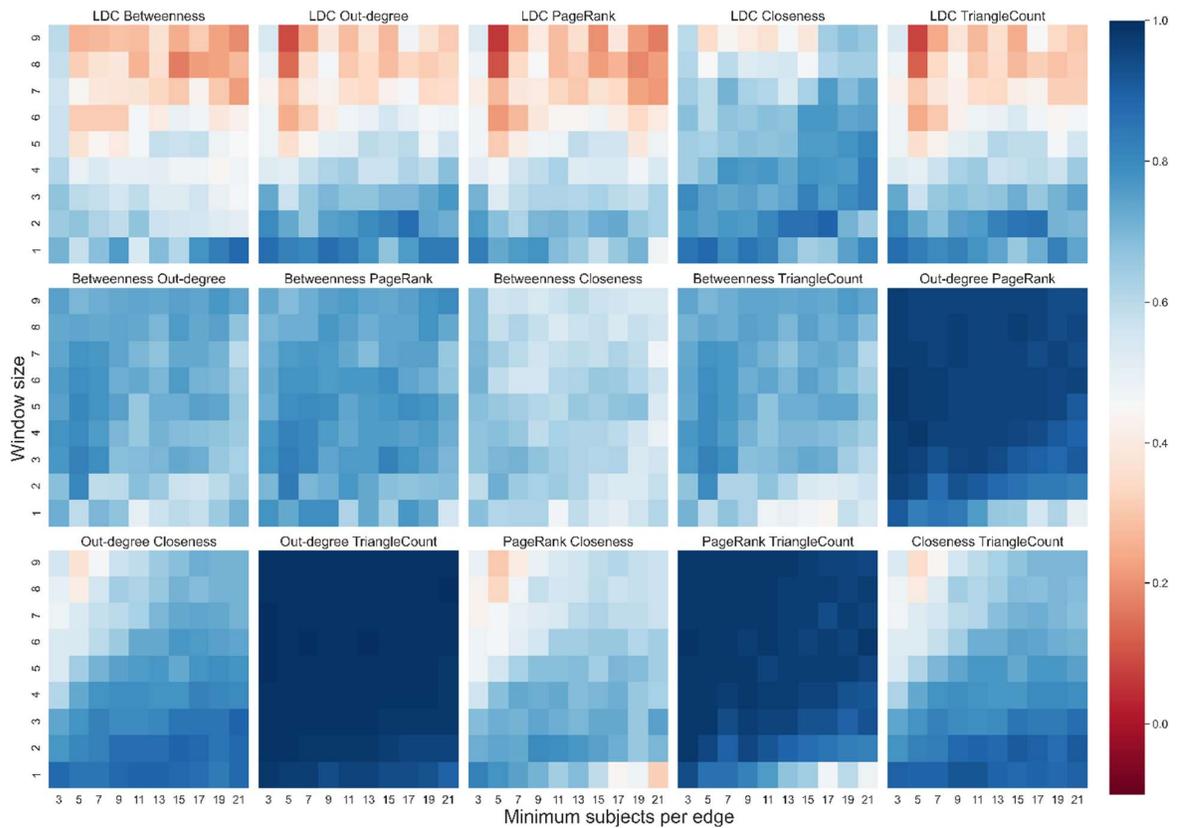

*Legend.* Each plot represents a correlation matrix between a pair of measures. The x-axis denotes the range of MS values, and the y- axis denotes the range of WS values. The color highlights the sign and magnitude of Spearman coefficients. The redder a cell is, the closer the correlation is to zero. The bluer a cell is, the closer the correlation is to one.

As Figure 3 indicates, the correlations between LDC and the other centrality measures are relatively low. In particular, as the window size (WS) in the distance function increases, the correlation between LDC and the other measures decreases. This pattern appears only in the matrices that include LDC and is therefore unique to the correlation between LDC and the other measures.

We performed a cluster analysis to determine whether LDC is linked to the other measures and, if so, to which specific ones. In addition, we used an anomaly detection test to examine the extent to which LDC is different from the other measures. The cluster analysis and anomaly detection tests were performed via the Scikit-learn package in Python (Pedregosa et al. 2011).

In the cluster analysis, hierarchical clustering merged vertices together one at a time, in a series of sequential steps, to result in homogeneous clusters. The goal was to increase within-group homogeneity and between-group heterogeneity. An agglomerative hierarchical cluster algorithm defined each vertex as a cluster, and in each iteration, clusters were merged to create a more significant cluster, with vertices in the same cluster more similar and vertices in different clusters more dissimilar. All clusters were merged into a single cluster at the end of the process. In this analysis, no assumption was made regarding the number of clusters. The distances between groups were the arithmetic mean distances between all the clusters' vertices (i.e., mean linkage).

The input for the cluster analysis was based on the Spearman correlation matrix between the measures. The correlation distance between two points was calculated as $1 - |r|$ where r denotes the Spearman correlation between two points.

Overall we performed the analysis for a range of window size (WS); in other words, the distances were set for a range from 1 to 9 of maximum words between any pair of words. For a given WS value, we analyzed the two values of 11 and 13, closest to the median (12) of the parameter MS, the minimum number of subjects per edge. Because the two analyses yielded similar results, here we present only the analysis for MS = 13.

We performed the analysis on nine correlation matrices, as reflected in Figure 5. Next, we used dimensionality reduction as a preprocessing step to compute the cluster analysis via an agglomerative hierarchical cluster. We used PCA to reduce the dimensions. Out of five principal components (PCs), PC-1 and PC-II contributed 89.7% to 93.9% of total cumulative variability. Therefore, both PC-I and PC-II were the axes of the final matrix in which we performed the agglomerative hierarchical clustering using the average criterion, which takes the average of the distances of each observation of the two clusters. Next, we produced a dendrogram that visualizes the grouping history (see Figure 4). The threshold we chose was 0.4. In all cases, the number of clusters was two. The results indicate that the larger the window size is, the further LDC is from the other centrality measures. By contrast, out-degree, PageRank, and the number of triangles are separated by a shorter distance as the size of the window increases. In addition, we found that within the context of the existing measures, LDC and closeness are relatively similar to one other.

**Figure 4.** *Visualization of the hierarchical clustering*

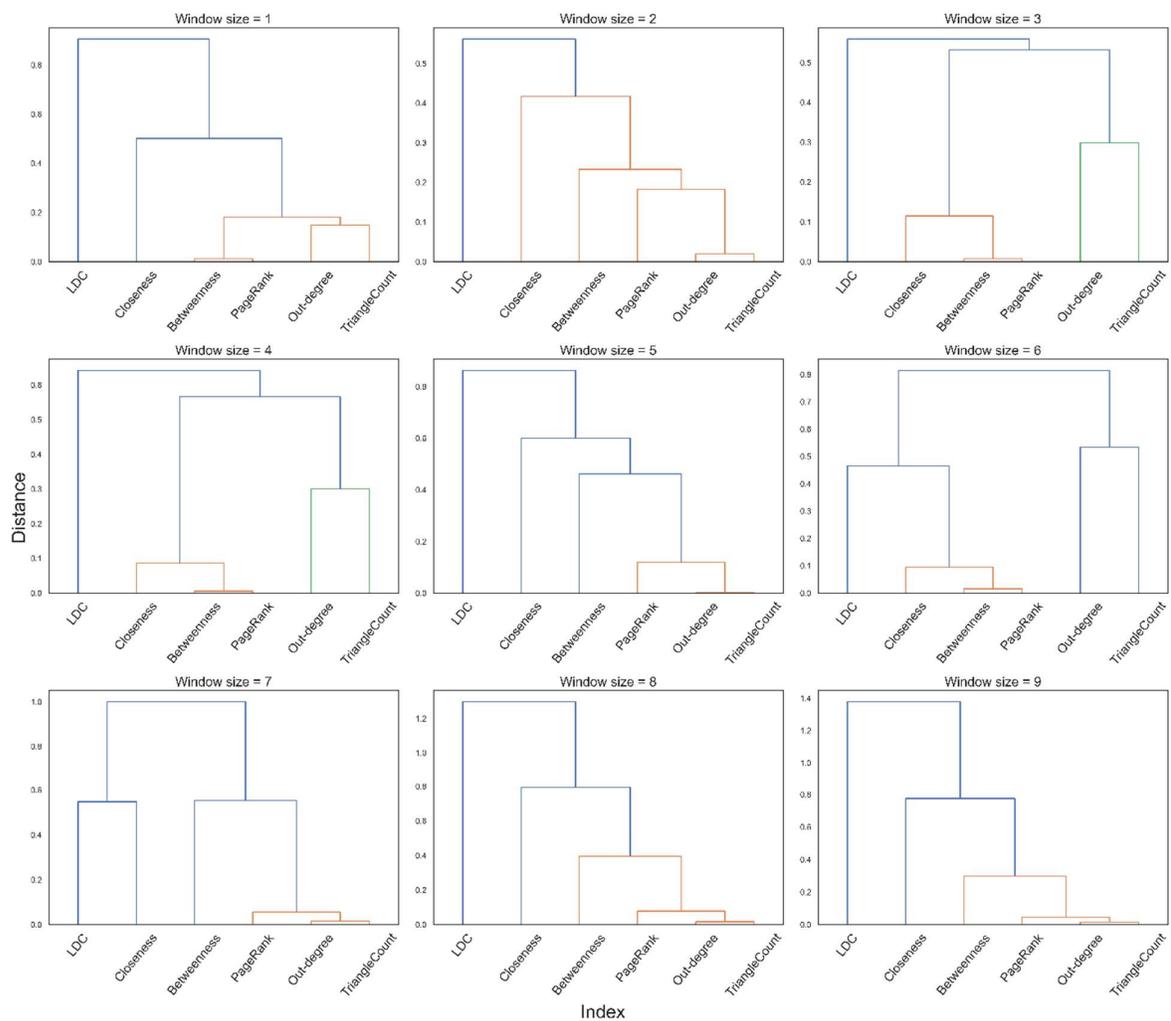

*Legend.* Each plot represents a dendrogram for a given WS value. MS was equal to 13 in all cases.

**Figure 5.** *Analysis on nine correlation matrices*

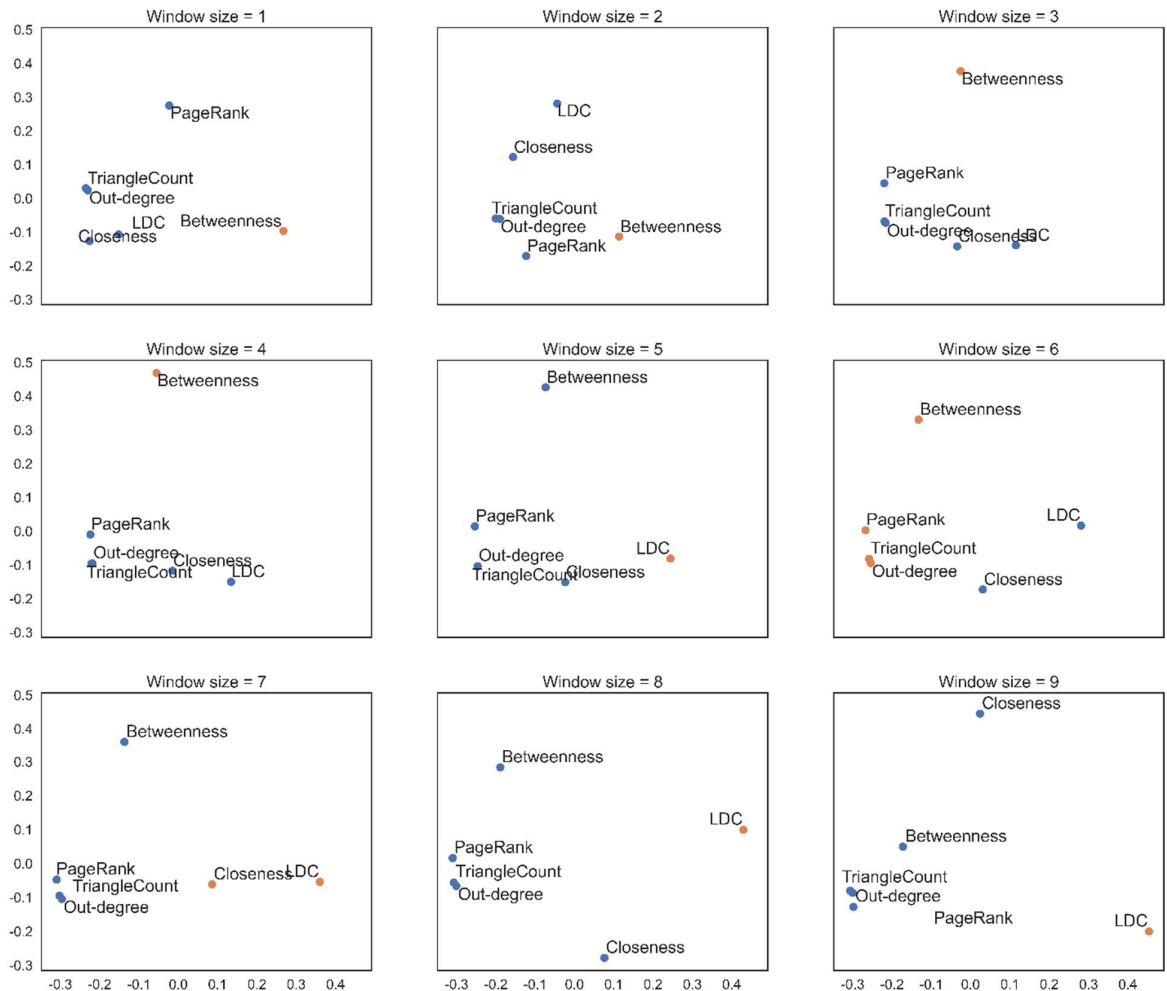

*Legend.* Each color denotes a different cluster. The x-axis and y-axis are PC-I and PC-II, respectively. Each plot represents a space for a given WS value, and MS was equal to 13 in all cases.

Finally, we examined the extent to which each measure deviates from the other measures. To this end, we ran IsolationForest, an unsupervised algorithm that assigns an anomaly score to each measure. The algorithm is based on a set of decision trees ($N = 100$) and assesses how the measure in question was isolated from the others. The closer the score of $p$ is to one, the more $p$ is different from other points. The closer the score of $p$ is to zero, the more likely $p$ is a normal point.

**Figure 6.** *IsolationForest to detect anomalies*

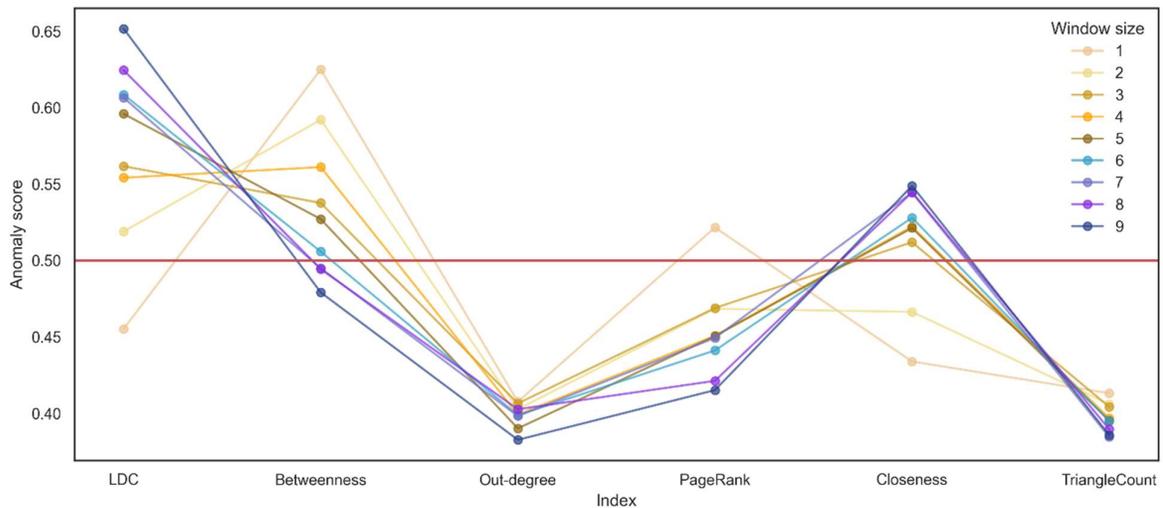

*Legend.* The x-axis represents different centrality measures, and the y-axis represents the anomaly score. Each line represents the anomaly score for any centrality measure for a given WS value.

As Figure 6 shows, LDC is above 0.5 in most cases. The greater the WS value is, the higher is the anomaly score of LDC. The opposite pattern can be found for betweenness: the lower the WS value is, the greater is the anomaly score of betweenness. In sum, an initial comparison indicates that LDC is different from some of the other measures. It can therefore be concluded that LDC captures a feature of centrality that is not fully expressed by closeness, betweenness, degree, number of triangles, and PageRank.

**Section 5. LDC as a measure of contextual diversity**

As we have already mentioned, long possible paths exist between groups of words that belong to different contexts, but the presence of a word that belongs to several contexts results in shorter possible paths between those contexts. The word that belongs to several contexts functions as an intermediary, and it may do so either because it belongs to a large number of semantic contexts or because it belongs to only a few contexts, each containing a large number of words. The novel measure presented in this study reflects the extent to which a word mediates between any pair of words in its vicinity. The more a word functions as an intermediary—in other words, the more paths that a word shortens on the graph—the greater the number of semantic contexts to which that word belongs. LDC takes into account more than the simple number of contexts in which a word appears. Instead, different contexts are weighted more heavily than similar contexts. A word that appears in distinct yet similar contexts will have a lower value than a word that belongs to very different contexts, with the number of contexts remaining constant.

Our understanding of centrality means that our novel measure must reflect more than the number of contexts in which a word appears, a measure suggested by Adelman et al. (2006), Brysbaert and New (2009), and McDonald and Shillock (2001). In our measure, different contexts are weighted more heavily than similar contexts. Given an equal number of contexts, we assign a lower value to a word that belongs to similar but distinct contexts than to a word that belongs to very different contexts. A similar approach, but not in regard to a semantic network, was suggested by Jones et al. (2012), who proposed the concept of semantic distinctiveness, the average dissimilarity across all of the documents in which the target word occurred. According to that study, semantic distinctiveness predicted more variance than CD and frequency in a word recognition task.

It has been suggested that *degree* can be used to identify CD words in a semantic graph. Degree consists of the number of edges that exist between a vertex, which represents a word, and other vertices (Hills et al. 2010; Sun and Pate 2017). But degree is highly correlated with frequency, as demonstrated in our study and others (Dorow et al. 2004), and frequency and CD are not necessarily the same phenomenon (for a review, see Caldwell-Harris 2021). For example, it has been found that CD words are recognized more quickly than words that are merely frequent, since words that belong to different contexts are more likely to appear in new contexts and are thus easier to access (Adelman et al. 2006). Additionally, since some edges may link words within similar contexts, degree—simply a count of the number of edges—does not necessarily capture the extent to which contexts are different. Our goal, then, is to offer a measure that expresses CD in a semantic graph that can be distinguished from frequency.

*Results*

As in the analysis we described in Section 4, we constructed 90 graphs, each one representing a different possible combination of the free parameters WS (window size), comprising the values [1,2,… ,9], and MS (minimum subjects per edge), comprising the values [3, 5, 7, 9, 11, 13, 15, 17, 19, 21]. We will start with the relationship between the centrality measures and frequency. Alternative centrality measures are highly correlated with log frequency: degree [M = 0.98, SD = 0.01], PageRank [M = 0.92, SD = 0.09], number of triangles [M = 0.98, SD = 0.01], closeness [M = 0.76, SD = 0.1], and betweenness [M = 0.74, SD = 0.07]). By contrast, the mean correlation between log frequency and LDC was 0.58 (SD = 0.15). Additionally, as we will show later in this paper, the larger WS and MS are, the weaker is the correlation between log frequency and LDC. This finding reinforces the claim that alternative centrality measures express frequency and do not adequately reflect CD.

**Figure 7.** *Correlation matrices between a centrality measure and log frequency*

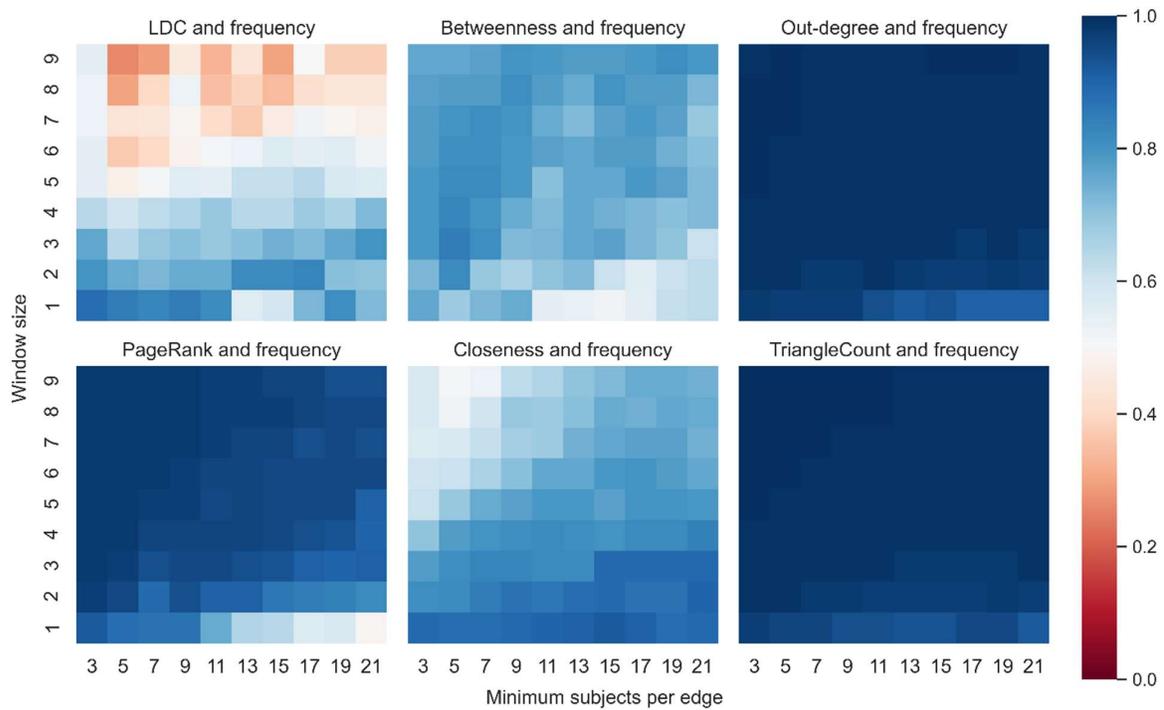

*Legend.* Each plot represents a correlation matrix between a centrality measure and log frequency. The x-axis denotes the range of MS values, and the y-axis denotes the range of WS values. The color highlights the sign and magnitude of Spearman coefficients. The redder a cell is, the closer the correlation is to zero. The bluer a cell is, the closer the correlation is to one.

In this section, we have shown that the alternative existing centrality measures are strongly correlated with frequency. More particularly, we found that degree, which was proposed as a measure for CD, is perfectly correlated with frequency. This finding is undesirable in light of the literature that emphasizes the need to distinguish between CD and frequency (Adelman et al. 2006; Caldwell-Harris 2021). In the next section, we will examine whether LDC also predicts accessibility to knowledge stored in memory when controlling for word frequency. Because the alternative centrality measures are highly correlated with frequency, we will not examine their performance.

## Section 6. Psychological validation: contextual diversity and accessibility

This section examines two questions. First, while the literature demonstrates that CD words are easier to access and therefore more easily recalled (Caldwell-Harris 2021), here we examine whether CD words as measured by LDC, are easier to access in a semantic memory task. In other words, are CD words recalled faster than non-CD words? Second, we examine whether CD words facilitate the retrieval of the words that follow it. Is the retrieval of a word that follows a CD word faster than the retrieval of a word that follows a non-CD word?

*Measure*

For any vertex $v \in V$ and all the paths $p \in P$, let $U$ be the set of paths $p_i, \ldots, p_n$ where $v_l^p$ is the l-th vertex in the path $p$. We calculate the average time (not normalized) that it takes to move from/to $v$ in $U$ as follows:

$$dt(v_{to}) = \frac{1}{|U|} \sum_{p \in P} \sum_{v_l = v \in p} t(v_l^p) - t(v_{l-1}^p) \tag{5}$$

$$dt(v_{from}) = \frac{1}{|U|} \sum_{p \in P} \sum_{v_l = v \in p} t(v_l^p) - t(v_{l+1}^p) \tag{6}$$

Two clarifications are critical. First, as suggested by Nachshon, Cohen, and Maril (2022), we distinguish between possible and actual paths. LDC is constructed by possible paths, meaning paths that can be drawn on the graph by connecting vertices for which the distance is defined. By contrast, dt-from/dt-to are constructed by actual paths—in other words, the list of words that the participants in the task actually retrieved.

Second, to rule out the possibility that the correlation between LDC and df-from/df-to is trivial—namely, that both measures are highly dependent on the same information—we distinguish between actual paths and random paths, which are shuffled lists of words from the actual paths. Note that the random paths preserve the frequency and number of words of each actual path. Our goal is to show that the correlation between LDC and dt-from/dt-to is significantly higher for the participants' actual word sequences than for the random paths. This means that the correlation between LDC and dt-from/to results from the order of the words in the actual paths.

*Results*

Spearman correlation was calculated between LDC and dt-from as well as between LDC and dt-to. The correlation was calculated for a range of parameters of the distance function: window size (WS), meaning the maximum number of words between a given pair of words that constitute an edge [1, 2,…,9], and minimum number of subjects (MS), meaning the minimum number of subjects per edge [3,5,7,9,11,13,15,17,19,21]. Of 90 correlations between LDC and dt-to, 82.2% were significant. For LDC and dt-from, 100% of the correlations were significant (see Figure 9). Figure 8 demonstrates some examples of the correlation between LDC and dt-from for a set of parameters WS: [2,7] and MS: [7,11,15].

**Figure 8.** *Six examples of the correlation between LDC and dt-from for WS: 2,7 and MS: 7,11,15*

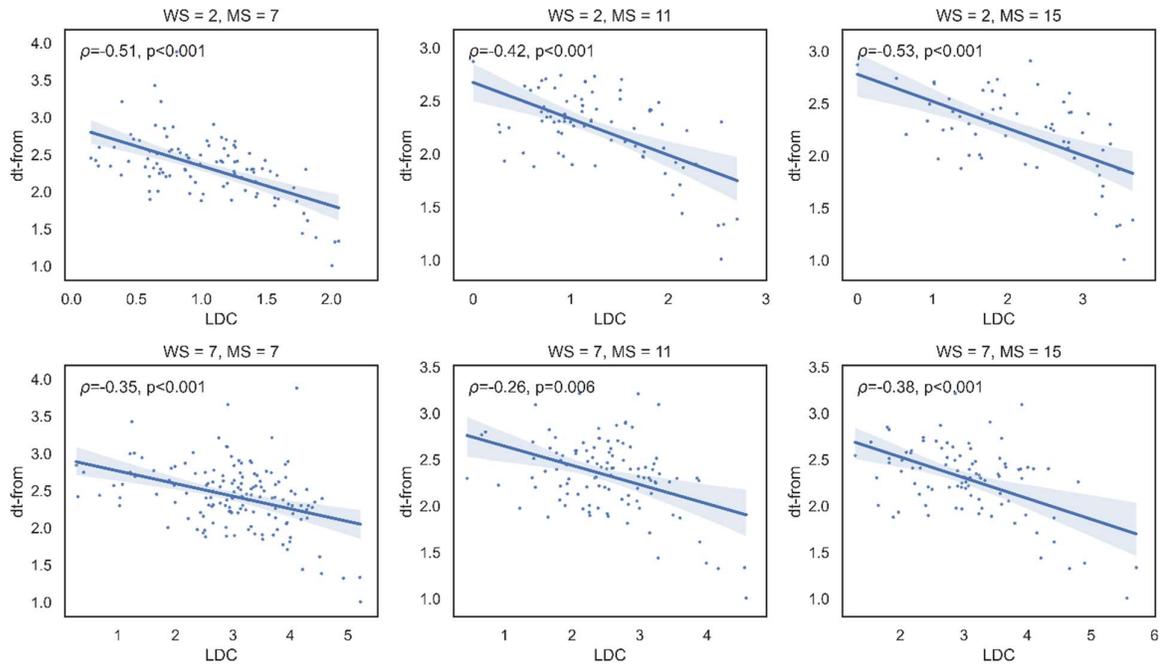

*Legend.* The x-axis denotes LDC and the y-axis denotes dt-from.

*Testing Triviality*

This section examines whether the correlations between LDC and dt-from/dt-to reflect a cognitive phenomenon rather than a trivial one. LDC and dt-from/to are highly dependent on the same information; our goal is to show that the correlations are significant only for lists of words generated by real participants.

To examine this issue, we distinguish between actual and random order. Actual order refers to the lists of words generated by real participants; random order refers to lists of words whose order is random.

Information about actual order is essential to the investigation of psychological phenomena such as retrieval. A person who retrieves a list of words is not providing a random sample from the distribution of word frequencies. Instead, the order of the words plays an essential role in free recall (Mandler and Dean 1969) and semantic models (Jones and Mewhort 2007). In our case, information about order is a fundamental feature in the semantic distance function, since the distances are determined by the amount of time it takes to pass from one word to another in an actual path. We expected the relationship between LDC and dt-from/dt-to to appear only in the participants' actual lists of words and not in random lists; if the relationship appeared in random lists, too, it would reflect a trivial connection between LDC and dt-from/dt-to.

We began by estimating the correlations between LDC and dt-from/ dt-to for a random order. Let $T$ be a set of paths and $t_S \in T$ be a path of subject $s$, where $t_S$ contains sequence word $t_{S1}$ to $t_{Sl_s}$ and the length of $t_S$ is $l_s$. First, let $T^r$ be a set of $t_S^r$ where $t_S^r$ is a random order of the path $t$ of subject $s$. Second, let $G^r = (V, E)$, such that the distances between the vertices are defined by $T^r$. Third, based on $G^r, \hat{G}^r$ and $T^r$, calculate $LWC^r$ and dt-from$^r$ /dt-to$^r$. Fourth, let $\rho^r$ denote the Spearman correlation between LDC$^r$ and dt-from$^r$ or LDC$^r$ and dt-to$^r$. The correlation $\rho^r$ can be generated by repeating steps one through four $N = 5000$ times, and as a result $P^r = (\rho_1^r, \dots, \rho_{100}^r)$ denotes the set of correlations obtained from the random orders. Finally, the distribution of $P^r$ defines the null hypothesis, and the statistical significance is the probability of obtaining the real correlation, obtained from the actual order, which is at most 5% at the null hypothesis.

The triviality was tested for any significant correlation between LDC and dt-from/dt-to given the following range of parameters of the distance function: WS: [1, 2,…,9]and MS: [3,5,7,9,11,13,15,17,19,21]. In total, 27 out of 90 cases were significant and not trivial; out of 34 significant correlations between LDC and dt-to, 79.4% were not trivial. For LDC and dt-from, out of 89/90 significant correlations between LDC and dt-to, 100% were not trivial (see Figure 9).

**Figure 9.** *Triviality test for the correlation between LDC and dt-to and between LDC and dt-from*

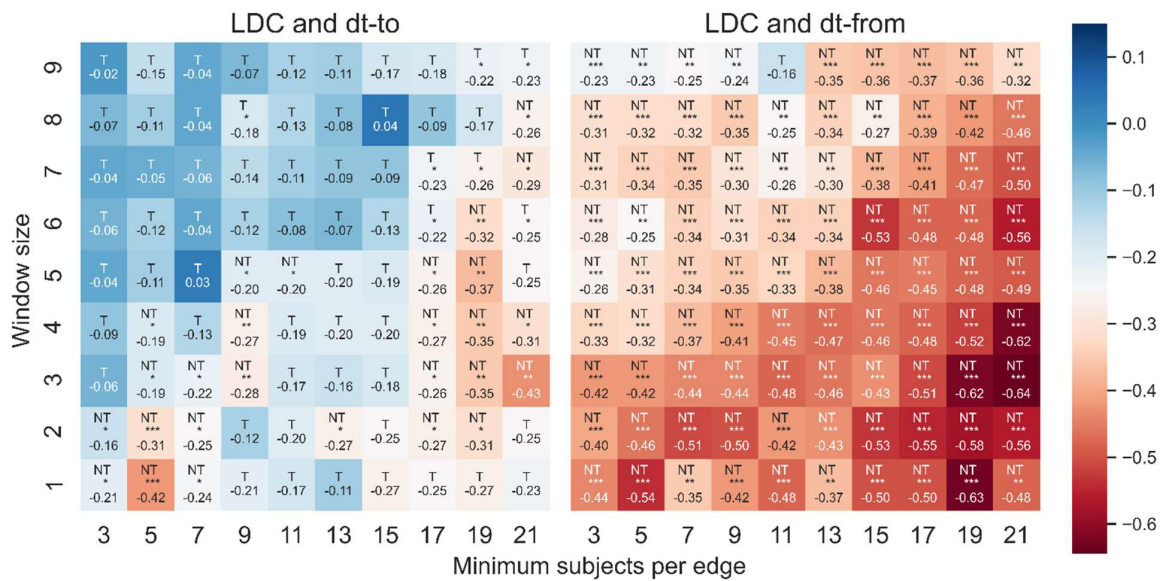

*Legend.* The color highlights the sign and magnitude of Spearman coefficients. The bluer a cell is, the closer the correlation is to zero. By contrast, red indicates negative values. The number of stars indicates the level of significance. NT/T denotes whether the correlation is non-trivial/trivial.

### Controlling for Frequency and Word Location

Finally, we examined whether LDC predicts dt-from/dt-to even when controlling for the frequency of the word and the average location of the word in the lists that people produced. To this aim, we ran a robust linear regression for all cases (i.e., pair of parameters MS and WS) where the correlation between LDC and dt-from/dt-to was not trivial.

Many studies have shown that word frequency is a strong predictor for how quickly a word can be named (Forster and Chambers 1973) as well as for lexical decision-making (Scarborough et al. 1977), perceptual identification (Morton 1969), and recall (Brysbaert and New 2009). We found as well that for the set of words with a frequency higher than 30 ($N$ = 132), there is a consistent and significant correlation between log-frequency and df-from and between log-frequency and df-to (see Figure 11).

In addition, we controlled the average location of the word, which maintains a positive correlation with dt-from and dt-to (see Figure 10). As the participant progresses further into the minute-long retrieval exercise, the transition time between consecutive words increases, and as a result, words that tend to appear at the beginning of the list are characterized by a higher retrieval speed. By controlling word location, we can examine whether the effect of LDC on dt-from/dt-to exists even when the location is taken into account, and we can also see whether the effect of LDC with dt-from/dt-to is dependent on word location. In particular, we can examine whether a CD word facilitates retrieval even as the stream of associations progresses and the subject has difficulty retrieving additional words, or whether the effect of a CD word depends on the word's location.

**Figure 10.** *Spearman correlation between log-frequency, the average location of a word, dt-from, and dt-to*

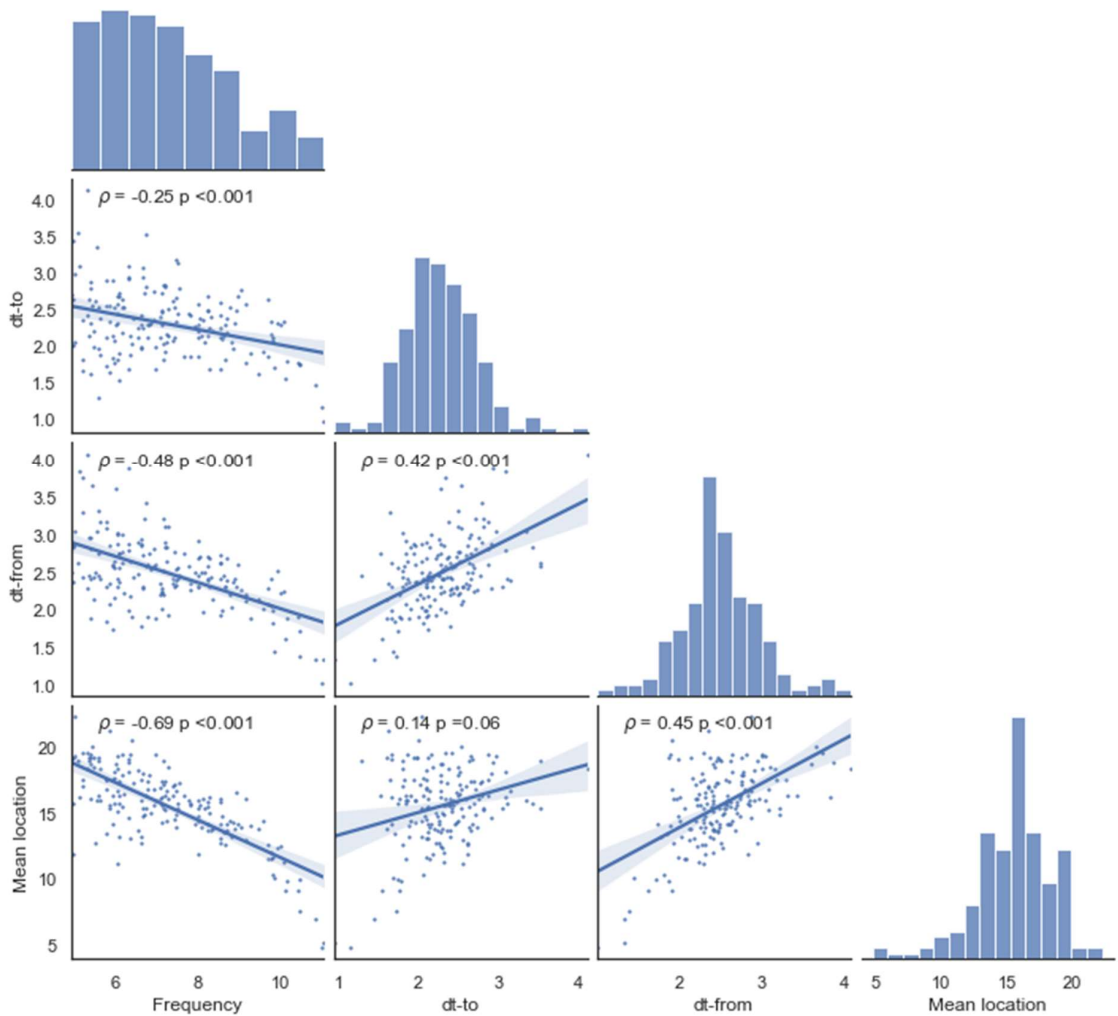

Using the statsmodels package in Python (Seabold and Perktold 2010), we analyzed the results with two robust linear regressions to assess the effect of LDC (continuous), log-frequency (continuous), and word average location (continuous) on dt-to. To assess the effect on dt-from, we used a second model, as follows: $y_i = \alpha + \beta_1 x_1 * \beta_2 x_2 * \beta_3 x_3 + \varepsilon$, where $Y_i$ is the independent variable dt-to or dt-from for word i, $\beta_1$ represents the fixed effect LDC, $\beta_2$ represents the fixed effect log-frequency, $\beta_3$ represents the fixed effect mean location, and $\varepsilon$ represents the residuals. All possible interactions were taken into account. Additionally,

LDC, log-frequency, and average locations were standardized via Z-score. We ran each model for any significant correlation between LDC and dt-from/dt-to given the following range of parameters of the distance function: WS:[1,2,…,9] and MS:[3,5,7,9,11,13,15,17,19,21].

As a first step, we tested multicollinearity by computing the variance inflation factor (VIF). Since log-frequency and average location are highly correlated, the mean VIF of log-frequency was above 2.5 (M = 3.56 25% = 2.72,  75% = 4.15), as was the mean VIF of average location (M = 3.85 25% = 2.59,  75% = 5.15). The average score of LDC was 2.19 (25% = 1.49,  75% = 2.8). For more details, see Figure A1 in the Appendix. We therefore broke each model into two: in the first one, we removed log-frequency, and in the second one, we removed the average location. In total, we ran four models, two to predict dt-to and two to predict dt-from.

These are the results for the dt-to models. For the first model (independent variables: LDC and log-frequency), out of 27 non-trivial and significant correlations between LDC and dt-to, only two betas were significant. For the second model (independent variables: LDC and average location), out of 27 non-trivial and significant correlations between LDC and dt-to, only one beta was significant (see Figure 12).

Next are the results from the dt-from models. For the first model (independent variables: LDC and log-frequency), out of 89 non-trivial and significant correlations between LDC and dt-from, 66 betas were significant. Naturally, the lower WS is, the lower is the number of vertices in the graph (see Figure 1). When the cases in which WS = 1,2 were removed, 59 out of 70 were significant (see Figure 12). For the second model (independent variables: LDC and log-frequency), 36 betas were significant. Note that the results are significant for the set of graphs with the highest number of vertices (see upper left corner of Figure 11).

**Figure 11.** *Beta coefficient between LDC and dt-to and between LDC and dt-from*

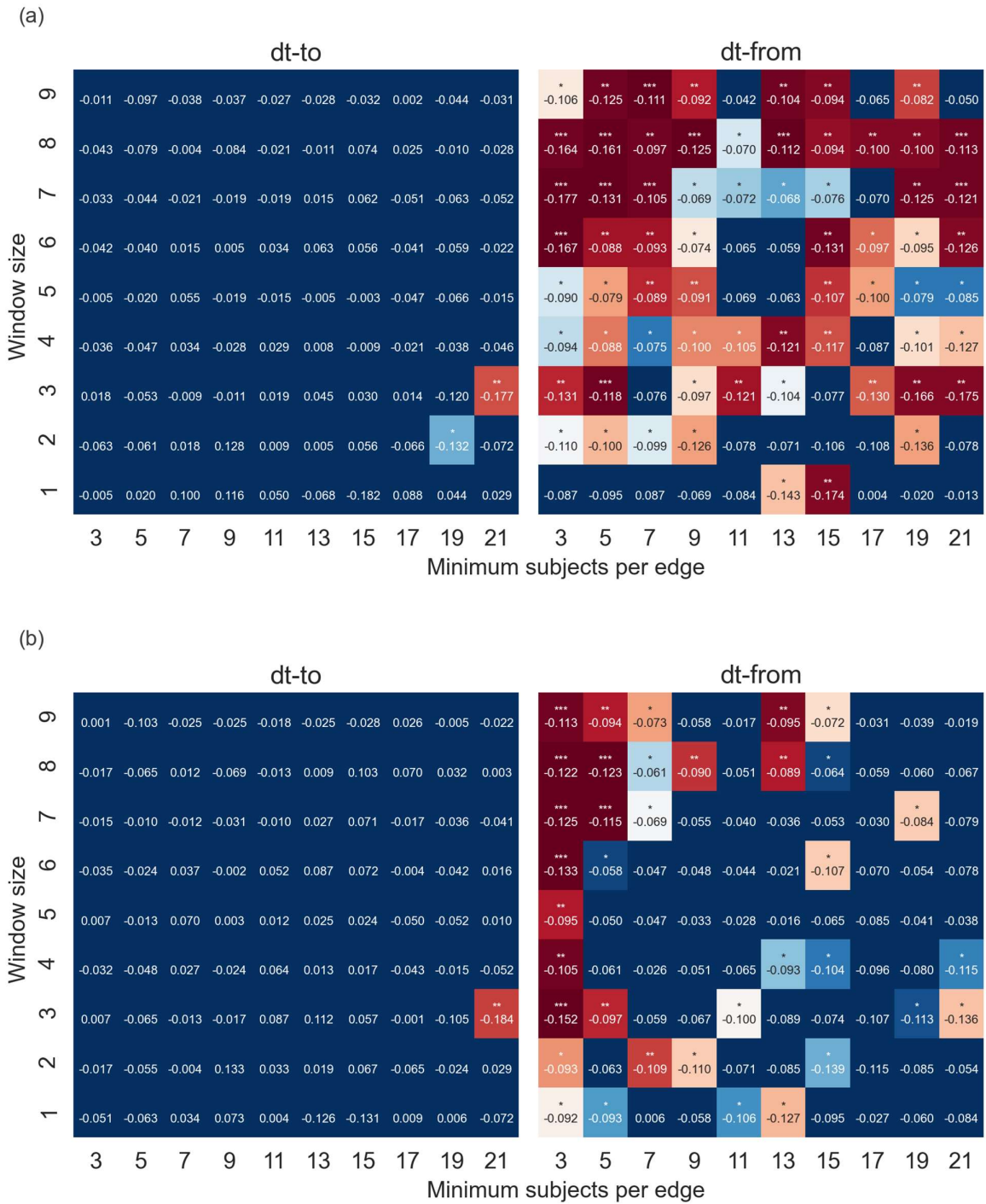

*Legend.* The color highlights the sign and magnitude of the beta coefficients between LDC and dt-from/dt-to. The blue cell indicates a non-significant beta. By contrast, the redder the cell is, the greater the beta coefficient is. The number of stars indicates the level of significance. The (a) graphs reflect the model with the independent variables LDC and log-frequency; the (b) graphs reflect the model with the independent variables LDC and average location.

Regarding the effect of the interaction between LDC and the average location on dt-to, a significant interaction was found for 26 out of 90 cases. The effect of the interaction on dt-from was robust; out of 90 cases, 89 were significant (Figure 12). In both cases—dt-from and dt-to—the effect of LDC depends on the word's location. In the first part of the stream of association (i.e., below the median location), LDC predicts dt-from/dt-to, and in the second part (i.e., above the median location), the relationship between LDC and dt-from/dt-to fades until it disappears (See Figures 13 and 14). Appendix 1 presents the beta coefficients of log-frequency, mean location, and log-frequency interaction with LDC.

To identify the regions in the moderator measure (average location) where the conditional effect of LDC on dt-from or dt-to were significant, we used the Johnson-Neyman floodlight analysis from the Python package PyProcessMacro (André 2021), a technique recommended by Spiller et al. (2013). We ran the floodlight analysis for every possible combination between WS and MS, a total of 90 cases. The direct effect of LDC on dt-to is on average significantly negative in 89 cases within the range of the interval values of the moderator average location, 4.93 (SD = 1.9) to 12.0 (SD = 2.36). The direct effect of LDC on dt-from is on average significantly negative in 90 cases within the range of the interval values of the moderator average location, 4.82 (SD = 0.62) to   14.08 (SD = 2.18).

**Figure 12.** *Interactions between location and LDC on dt-to and between location and LDC on dt-from*

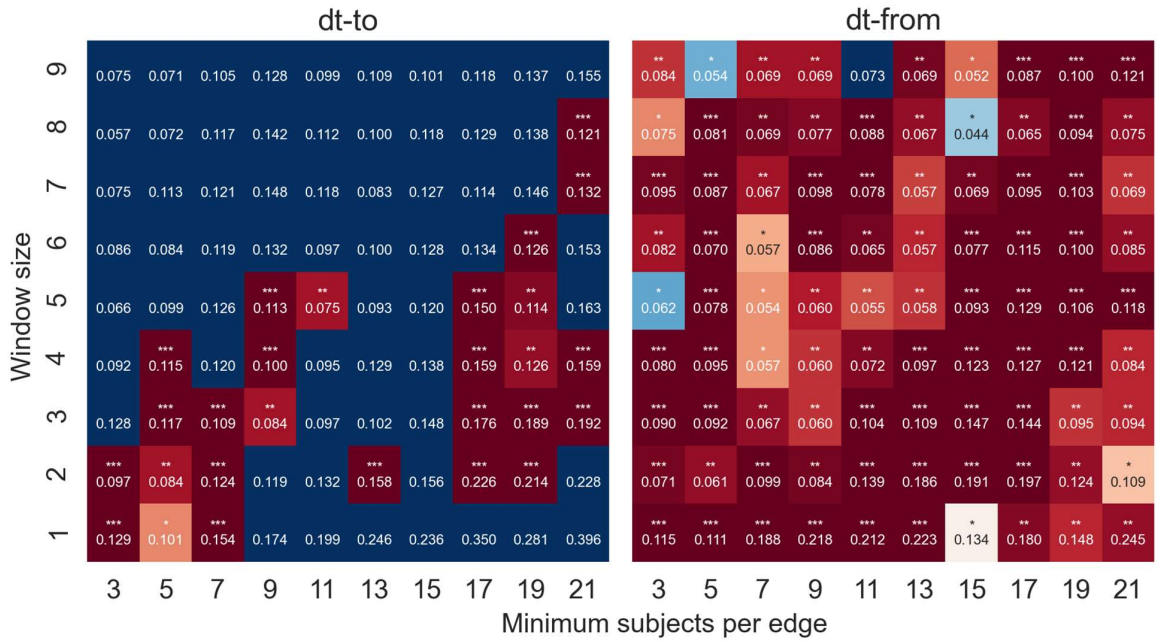

*Legend.* The color highlights the sign and magnitude of the beta coefficients. A blue cell indicates a non-significant beta. By contrast, the redder the cell is, the greater the beta coefficient is. The number of stars indicates the level of significance.

**Figure 13.** *Examples of interactions between LDC and average location on dt-to for WS: 2,4 and MS: 5,7,17*

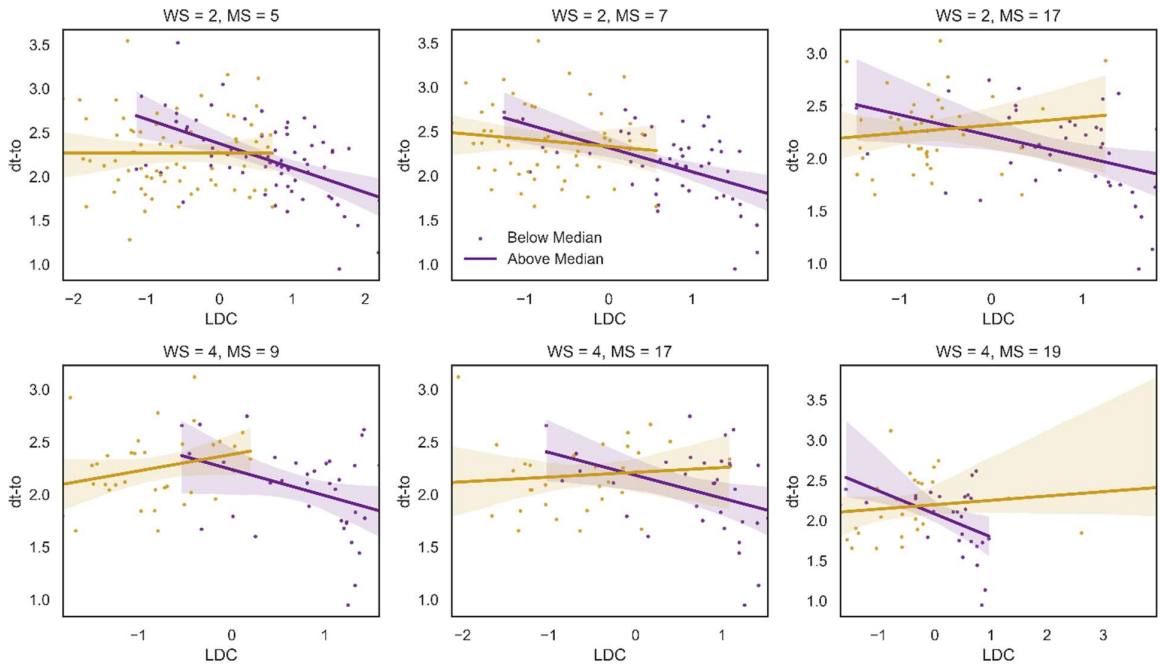

*Legend.* The x-axis denotes LDC, and the y-axis denotes dt-to.

**Figure 14.** *Examples of interactions between LDC and average location on dt-from for WS:*

*2,7 and MS: 7,11,15*

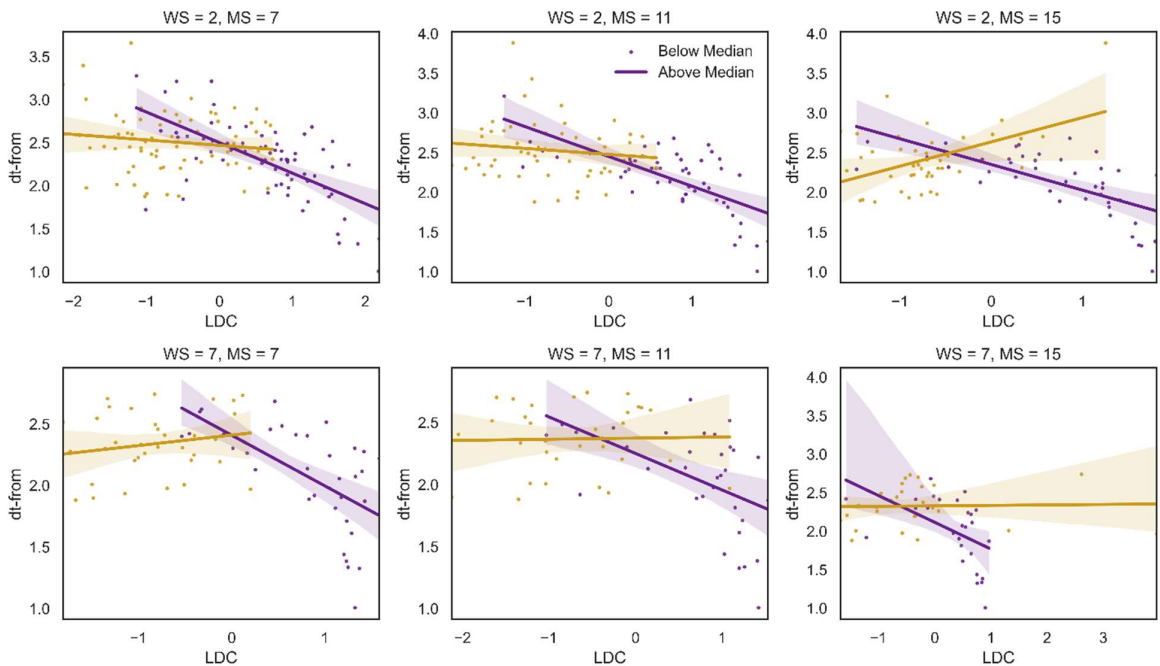

*Legend.* The x-axis denotes LDC and the y-axis denotes dt-from.

**Section 7. Discussion and Conclusion**

LDC measures to what extent a vertex functions as a local intermediator, or, in other words, to what extent a vertex shortens possible paths between neighboring vertices. To arrive at this measure, we essentially subtract two matrices, one that includes the shortest paths given the set of vertices that are in the vicinity of $v$ when it is possible to pass through $v$, and one that includes the shortest paths given the same set of vertices when it is not possible to pass through $v$. The higher the value of $v$ is, the more $v$ binds unrelated words in its local environment.

From a general perspective, controlling the flow of information in the graph means systematically offering a faster way of moving from one vertex to another than is possible with alternative paths. In this broad context, further research might compare centrality measures in an undirected network or compare the performance of LDC to other betweenness centrality measures such as random-walk betweenness and current flow betweenness. LDC might also be extended to include non-geodesic paths as does flow betweenness centrality. In this case, the comparison to alternative paths would be based not only on the shortest paths but on all paths passing through the vertex. In addition to its use as a means of locating central vertices, LDC can also be applied as a way of detecting communities; this application would involve removing edges that are connected to high betweenness centrality. Finally, an essential question that grows out of our work is the relationship between LDC and various notions of Ricci curvature for a network, such as Forman, Ollivier, Menger, and Haantjes. This question will need to constitute the subject of further research.

From the semantic perspective, this work is part of a study by Nachshon, Cohen, Naim et al. (2022), which attempts to characterize the relationship between cognitive possesses such as retrieval patterns and the structure of the semantic network. Here we have focused on the relationship between the psycholinguistic interpretation of our new measure as a means of capturing CD words, which is a network property, and accessibility to memory measured by df-from/df-to, which is a property of the cognitive process.

We have demonstrated that existing measures do not fully explain LDC because significant differences exist between LDC and other centrality-based relationships, namely out-degree, closeness, PageRank, and betweenness. While a centrality-based relationship such as degree maintains an almost perfect relationship with frequency, LDC's relationship with frequency is weaker. These findings reinforce the claim that degree is not a good approximation for contextual diversity in a semantic graph.

Next, we offer a psychological validation of LDC by examining two ways in which a CD word affects retrieval processes in a serial semantic task. On the one hand, based on previous literature that shows high accessibility of CD words (Adelman et al. 2006; Brysbaert and New 2009; Johns et al. 2012; Baayen 2010; Caldwell-Harris 2021; Steyvers and Malmberg 2003), we expected faster transitions to a high CD word in our study as well. On the other hand, a previous study suggested that the high accessibility found in non-serial tasks may not be generalizable to all retrieval tasks, and specifically that high accessibility is not observed in a serial recall, or episodic, task (Guitard et al. 2019). These latter findings weakened our expectation of finding significant results regarding the effect of LDC on dt-to. The results mostly indicate a non-significant and/or trivial relationship between LDC and dt-to when controlling for frequency or average location. However, the interaction between LDC and average location yields better results in predicting dt-to than when LDC is considered alone. As demonstrated by floodlight analysis, the interaction between LDC and average location indicates that the negative relationship between CD and retrieval speed appears primarily at the beginning of a serial task. Therefore, although it is inconclusive on this subject, our study is consistent with previous findings that significant results in a non-serial task do not necessarily appear in a serial task (Guitard et al. 2019).

On the question of whether CD words facilitate retrieval of the upcoming word, we found that the correlation between LDC and dt-from is significant and not trivial, particularly when we controlled for log-frequency. When we controlled for average location, the result was weaker but significant for the set of graphs with the highest number of vertices. Additionally, the interaction between average location and LDC leads to a robust effect on dt-from. The effect of LDC on dt-from is apparent at the beginning of the stream of associations and weakens over time as approximated by average location. This finding highlights the difference between serial tasks, which involve an ongoing retrieval process, and non-serial tasks, which involve other forms of retrieval. In a serial task, word location as a mediating factor in the relationship between CD and retrieval speed may reflect the importance of the subject's retrieval history. One possibility is that as the subject progresses in the task, the retrieval of the next word is conditioned by the retrieval history that preceded it, and the history masks the relation between CD and retrieval speed. Another factor may be the fatigue that is manifested in the difficulty of retrieving additional words as the task progresses; perhaps this fatigue masks the relationship between CD and retrieval speed as the subject progresses in the task. Further research may identify the factors that mediate the relationship between CD and declining retrieval speed as approximated by the average location.

**List of Abbreviations**

**CD:** contextual diversity

**LDC**: Local Detour Centrality

**MS:** minimum subjects

**WS:** window size

**Declarations**

*Availability of data and materials*

The datasets that were generated and/or analyzed in this study are available in the repository of the Open Science Framework, doi: 10.17605/OSF.IO/NQWU7.

*Competing interests*

The authors declare that they have no competing interests.

*Funding*

This research was partially supported by the German-Israeli Foundation (grant I-1514-304.6/2019 to Emil Saucan and Jürgen Jost) and the Israel Science Foundation (grant 1471/20 to Anat Maril).

*Authors' contributions*

Conceptualization, H.C., Y.N., P.N., A.M. and E.S.; methodology, H.C. and Y.N.; formal analysis, H.C., Y.N., P.N., J.Y. and E.S.; statistical analysis, H.C.; writing—original draft preparation, H.C.; writing—review and editing, H.C., J.Y. A.M. and E.S.; visualization, H.C.; supervision, A.M. and E.S.; All authors have read and agreed to the published version of the manuscript.

*Acknowledgments*

Not applicable

**Supplementary information: Appendix**

**Figure A1.** *Testing multicollinearity by computing the variance inflation factor of the predictor variables*

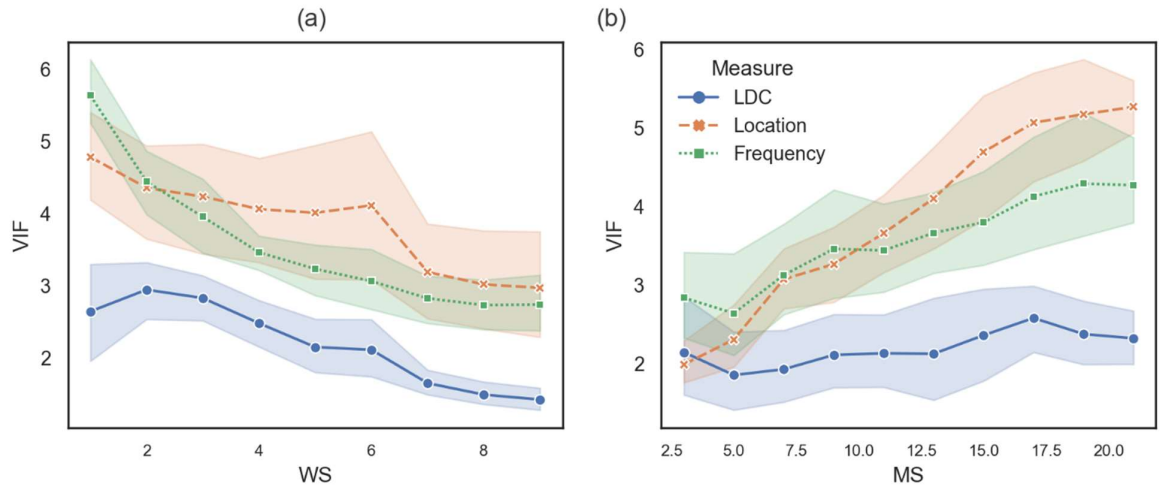

*Legend.* Part (a) presents the VIF score for each measure as a function of WS, while part (b) presents the VIF score for each measure as a function of MS.

**Figure A2.** *Correlation matrices between a centrality measure and average location*

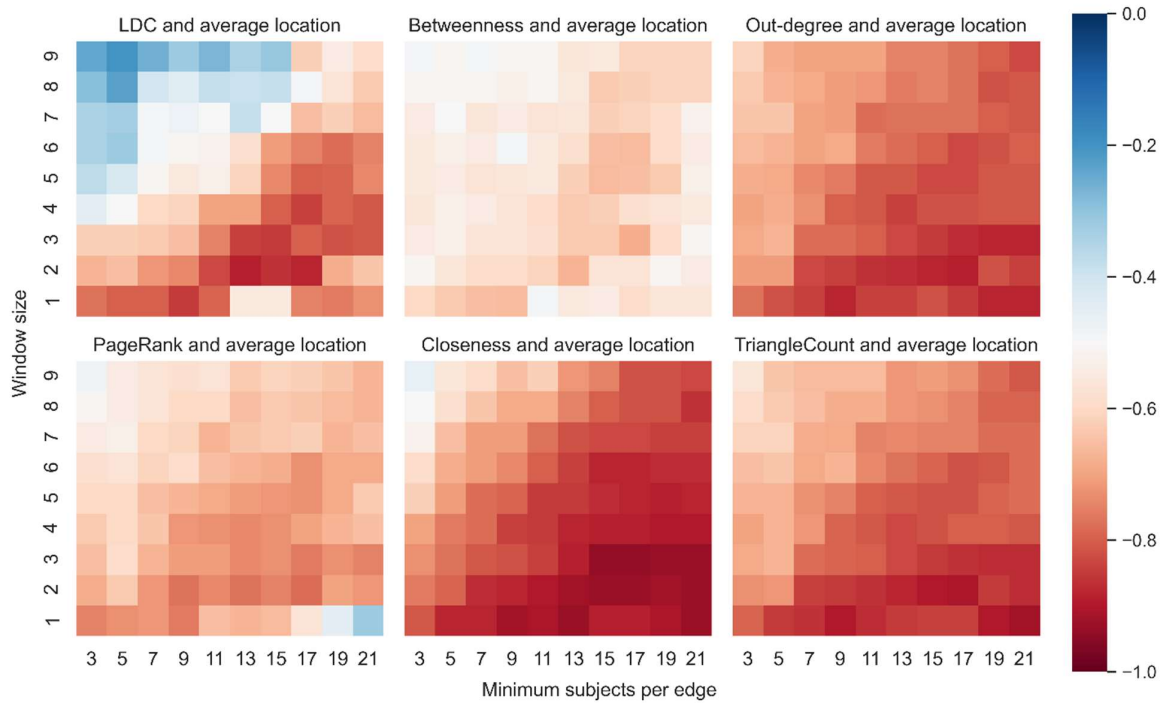

*Legend.* Each plot represents a correlation matrix between a centrality measure and average location. The x-axis denotes the range of MS values, and the y-axis denotes the range of WS values. The color highlights the sign and magnitude of Spearman coefficients. The redder a cell is, the closer the correlation is to zero. The bluer a cell is, the closer the correlation is to one. The mean correlation between average location and each centrality measure came out as follows: degree -0.77 (SD = 0.07), PageRank (M = -0.65, SD = 0.07), number of triangles (M = -0.8, SD = 0.08), closeness (M = -0.8, SD = 0.1), betweenness (M = -0.54, SD = 0.04), and LDC (-0.59, SD = 0.18).

**Figure A3**. *Beta coefficients between frequency and dt-to/dt-from*

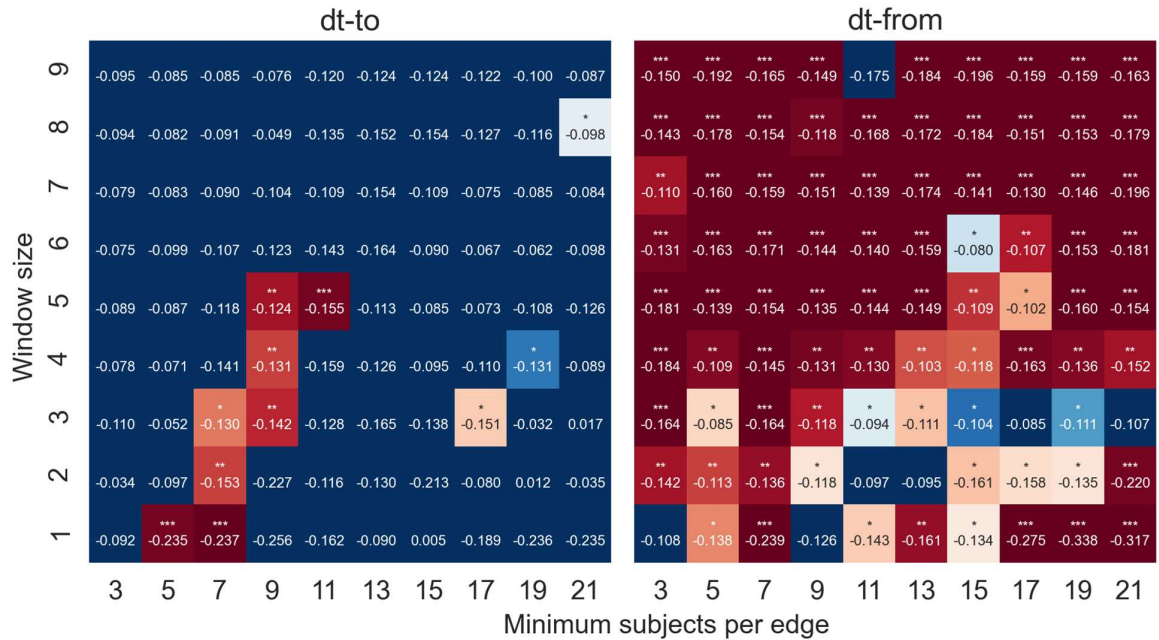

*Legend.* The color highlights the sign and magnitude of the beta coefficients. The blue cell indicates a non-significant beta. By contrast, the redder the cell is, the greater is the beta coefficient. The number of stars indicates the level of significance.

**Figure A4.** *Beta coefficients between the interaction of LDC and frequency, and dt-to/dt-from*

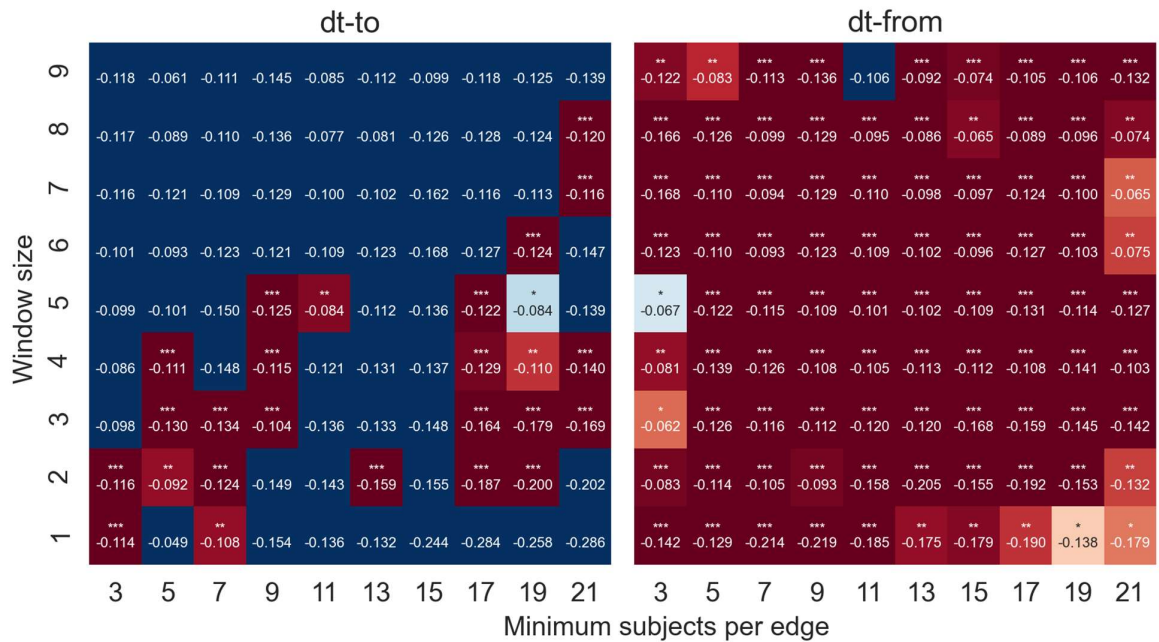

*Legend.* The color highlights the sign and magnitude of the beta coefficients. A blue cell indicates a non-significant beta. By contrast, the redder the cell is, the greater is the beta coefficient. The number of stars indicates the level of significance.

**Figure A5.** *Beta coefficients between average location and dt-to/dt-from*

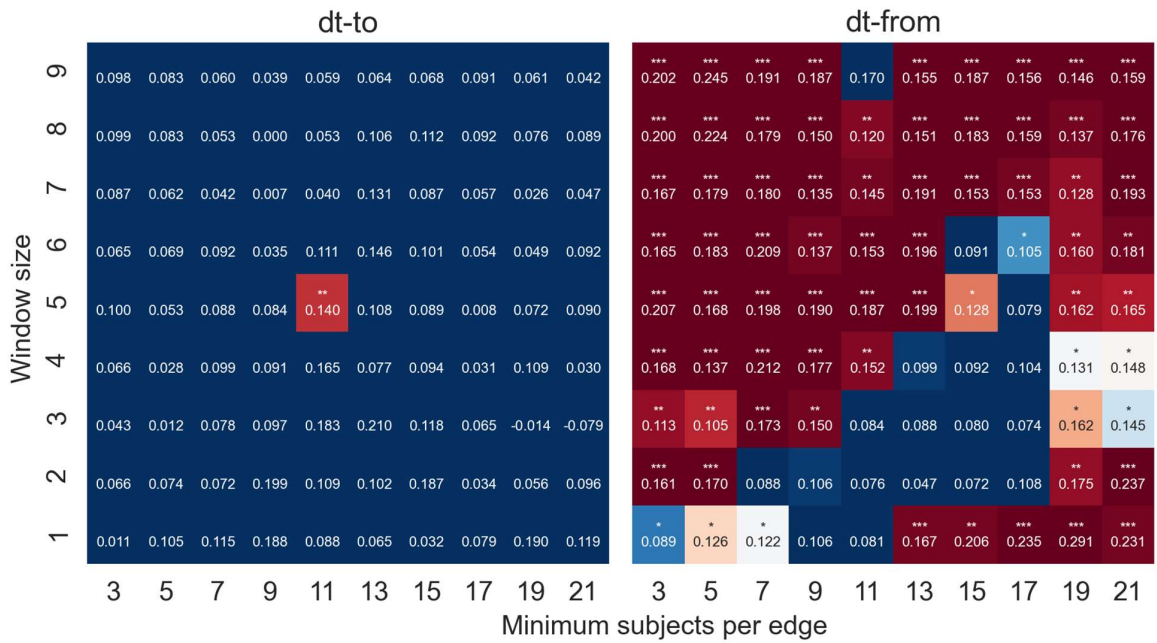

*Legend.* The color highlights the sign and magnitude of the beta coefficients. A blue cell indicates a non-significant beta. By contrast, the redder the cell is, the greater is the beta coefficient. The number of stars indicates the level of significance.

**Figure A6**

*Beta coefficients between LDC and dt-to/dt-from for regression model with predictors LDC and frequency*

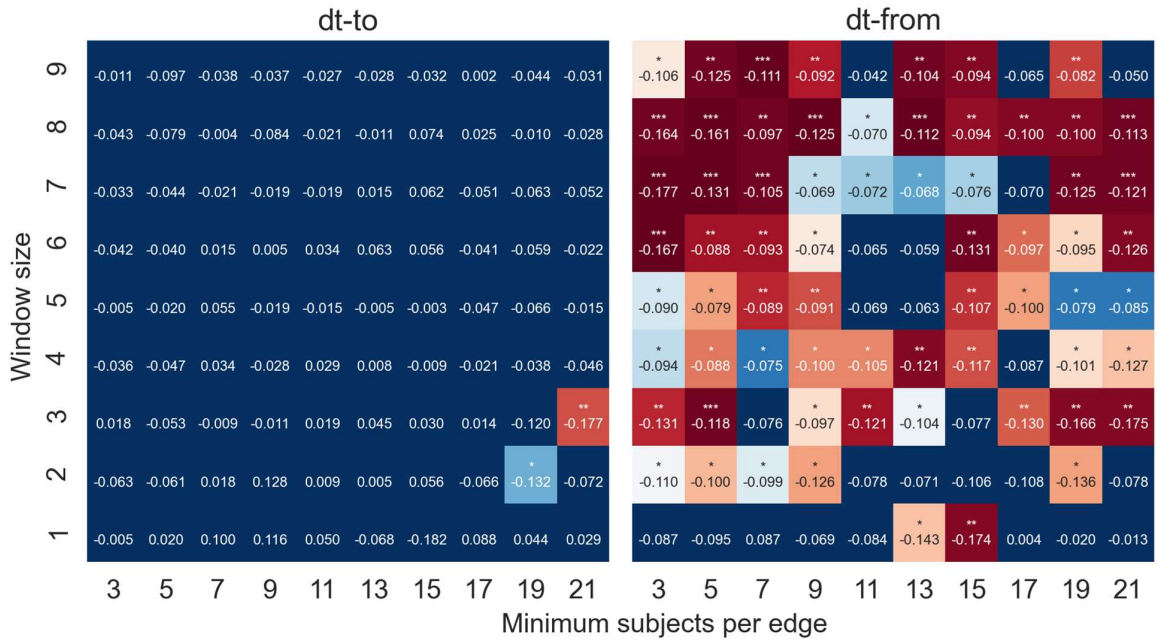

*Legend.* The color highlights the sign and magnitude of the beta coefficients. A blue cell indicates a non-significant beta. By contrast, the redder the cell is, the greater is the beta coefficient. The number of stars indicates the level of significance.